\documentclass[12pt]{article}
\usepackage{epsfig}
\newcommand{\be}{\begin{eqnarray}}
\newcommand{\ee}{\end{eqnarray}}

\newcommand{\AmS}{{\protect\the\textfont2
  A\kern-.1667em\lower.5ex\hbox{M}\kern-.125emS}}
\newlength{\dinwidth} \newlength{\dinmargin}
\begin{document}

\begin{flushright}
  Cavendish-HEP-02/16\\
  NIKHEF/2002-014\\
  ITP-2002/62\\
  DESY 02-216\\
  LBNL-51590\\
  \today\\
\end{flushright}

\vspace{8mm}
\begin{center}
  {\Large\bf\sc Threshold Effects in Charm Hadroproduction}
\end{center}

\vspace{0.5cm}
\begin{center}
  {\large Nikolaos Kidonakis$^a$, Eric Laenen$^b$, Sven Moch$^{c}$, 
Ramona Vogt$
^d$}\\
  \vspace{6mm}
  $^a${\it Cavendish Laboratory \\
    University of Cambridge, Cambridge CB3 0HE, UK} \\
  \vspace{2.5mm}
  $^b${\it NIKHEF Theory Group\\
    P.O. Box 41882, 1009 DB Amsterdam, The Netherlands \\
  and \\
Institute for Theoretical Physics\\ Utrecht University,
   Utrecht, The Netherlands}\\
  \vspace{2.5mm}
  $^c${\it Deutsches Elektronensynchrotron \\
   DESY Platanenallee 6, D--15738 Zeuthen, Germany} \\
  \vspace{2.5mm}
  $^d${\it Nuclear Science Division,\\
    Lawrence Berkeley National Laboratory, Berkeley, CA 94720, USA \\
  and \\
  Physics Department,\\
    University of California at Davis, Davis, CA 95616, USA}
\end{center}
\vspace{0.5cm}

\begin{abstract}
We describe calculations of $c \overline c$
production to next-to-next-to-leading order (NNLO) and next-to-next-to-leading
logarithm (NNLL) near threshold in $pp$ and $\pi^- p$ interactions.
We study the relevance of these calculations for existing $c \overline c$
total cross section data by examining their sensitivity to partonic
threshold kinematics, their convergence properties and scale dependence.
\end{abstract} 
\thispagestyle{empty} \newpage \setcounter{page}{2}

\section{Introduction}\label{sec:introduction}

Calculations of charm production are still not under solid
theoretical control.  A good understanding of the charm cross section is of 
particular interest for heavy ion physics.  Charm production is an important 
contribution to the dilepton continuum in heavy ion collisions.  In S+U and 
Pb+Pb interactions at the CERN SPS, with $\sqrt{S} = 19.4$ and 17.3 GeV/nucleon
pair\footnote{Unless otherwise specified, all $pA$ energies
are per nucleon and nucleus-nucleus energies are per nucleon pair.}
 respectively, the dilepton yield in 
the mass range $1.5 < m_{\mu^+ \mu^-} < 2.5$ GeV is enhanced by a factor of 
2-3 over the extrapolated proton-nucleus, $pA$, yield
\cite{na50}.  These data have been interpreted as an enhancement of the 
charm production cross section in the system created in the heavy ion 
collision. Another possible source of the dilepton enhancement is thermal 
dilepton production, predicted in a quark-gluon plasma \cite{grrapp}.  
Finally, the total charm rate would be
a useful reference for $J/\psi$ production in heavy ion collisions
\cite{grrapp,goren12,goren3}.

Although many future heavy ion experiments will be at high collider energies,
$\sqrt{S} \geq 130$ GeV, 
some of the current and future experiments, like those
at the SPS, are in the near-threshold region.  The NA60
experiment was specifically designed to distinguish between charm decays and
Drell-Yan-like production of dileptons \cite{na60} to determine the cause of
the apparent enhancement.  It will take heavy ion data at $\sqrt{S} = 17.3$
GeV and $pA$ data at $\sqrt{S} = 29.1$ GeV.  
A new facility is being built at
the GSI \cite{gsi} that will measure charm near threshold with $\sqrt{S}=6.98$
GeV.

Because the charm quark mass is a few times $\Lambda_{\rm QCD}$, 
it is generally treated as a
heavy quark in perturbative QCD calculations.  However, its relative
lightness results in a rather
strong dependence of the total cross section on mass and scale, with up to a
factor of 100 between the lowest and highest next-to-leading order (NLO)
results \cite{mlm}.  There is also a rather broad spread in the measured charm 
production cross section data at fixed target energies.  
Much of this uncertainty arises from
low statistics in the early experiments, assumptions of how much of the total
charm yield results in final-state $D$ mesons, and how the measured data are
extrapolated to full phase space.
The more recent data have improved considerably with new detection techniques
and higher statistics.

Improvements in the calculation of the charm cross section are difficult at 
all energies, but are perhaps possible when the $c \overline c$ pair is 
produced close to threshold, as we now describe. 
Factorization properties of QCD separate cross sections into
universal, nonperturbative parton densities and a perturbatively calculable
hard scattering function, the partonic cross section.  Remnants
of long-distance dynamics in the hard scattering function can dominate
corrections at higher orders near production threshold.  These Sudakov
corrections have the form of distributions singular at partonic threshold.
Threshold resummation techniques organize these singular distributions to all
orders, presumably extending the reach of QCD into near-threshold production. 
 The singular functions organized by resummation are plus distributions,
of the form $[\ln^l x/x]_+$, 
where $x$ denotes the `distance' from partonic threshold.
At next-to-leading log (NLL) accuracy and beyond, 
proper account must be taken of the color structure of 
the hard scattering \cite{KS,KOS} for each partonic subprocess.

Resummed cross sections are useful as generating functions for
approximate finite order corrections to the 
cross section when expanded in powers of the strong coupling constant
$\alpha_s$, as we do in this paper.
The resummed cross sections may also be evaluated numerically.
The charm fixed-target data were first compared to a leading log (LL)
resummed calculation of the total cross section in Ref. \cite{SVchm}.  
Because the ratio $m/\Lambda_3$ is quite small, the expansion parameter, 
$\alpha_s$, is not and the LL resummation began to fail at
$\sqrt{S} \approx 20$ GeV.  A NLL resummed evaluation in Ref. \cite{Bonci} 
found significant threshold corrections, albeit with a reduction in 
scale dependence.

In this paper we work at finite order, using our 
results of Refs.~\cite{NK,KLMV}. We have calculated the 
double-differential heavy
quark hadroproduction cross sections up to next-to-next-to-leading order
(NNLO), ${\cal O}(\alpha_s^4)$, and next-to-next-to-leading logarithm (NNLL),
{\it i.e.}\ keeping powers of the singular functions as low as $l = 2i - 1$ 
at order ${\cal O}(\alpha_s^{i+3})$ where $i=0,1, \ldots$ \cite{NK,KLMV}.
We only discuss $Q \overline Q$ production in the $ij = q \overline q$ and $gg$
channels since $qg$ scattering first appears at NLO.

Our studies focus on the kinematics of the proposed GSI facility and
the CERN SPS proton and ion fixed target programs.  
We first briefly describe our NNLO-NNLL calculations in the next section.  
In section 3 we discuss whether the ion beam energies will
actually produce charm in the near-threshold region where our calculations are
in principle applicable.  We show results for several values of the charm 
quark mass, $m= 1.2$, 1.5 and 1.8 GeV and for scales $\mu = m$ and $2m$.  
We compare our results for the NNLO-NNLL inclusive $c \overline c$
cross section to charm production data and to the NLO cross sections
in the relevant energy regime, and judge their value. 
Finally, we summarize our results in section 4.

\section{Resummation}\label{sec:resummation}

In our approach, the distance from partonic threshold in the singular
functions depends on how the cross section is calculated.  We either integrate
over the momentum of the unobserved heavy quark or antiquark and determine
the one-particle inclusive (1PI) cross section for the detected quark or
treat the $Q$ and $\overline Q$ as a pair in the integration, in pair 
invariant mass (PIM) kinematics.   In 1PI kinematics,
\begin{eqnarray}
\label{eq:3}
p(P_1) + p(P_2) \longrightarrow Q(p_1) + X(p_X)\, , 
\end{eqnarray}
where $X$ denotes any
hadronic final state containing the heavy antiquark and
$Q(p_1)$ is the identified heavy quark.  
The reaction in Eq.~(\ref{eq:3}) is dominated 
by the partonic reaction
\begin{eqnarray}
\label{eq:7}
i(k_1) + j(k_2) &\longrightarrow& Q(p_1) +
X[\overline Q](p_2')\, . 
\end{eqnarray}
At LO or if $X[\overline Q](p_2') \equiv \overline Q(\overline p_2)$, 
the reaction is
at partonic threshold with $\overline Q$ momentum $\overline p_2$.  
At threshold the
heavy quarks are not necessarily produced at rest but with
equal and opposite momentum.
The partonic Mandelstam invariants are
\begin{eqnarray}
  \label{eq:9}
s =(k_1+k_2)^2 \, ,\quad t_1 = (k_2-p_1)^2 -m^2 \, ,\quad
u_1 = (k_1-p_1)^2 -m^2 \, ,\quad s_4 = s + t_1 + u_1 \, 
\end{eqnarray}
where the last, $s_4 = (p_2')^2 - m^2$, is the inelasticity of the partonic
reaction.  At threshold, $s_4=0$.  Thus the distance from threshold in 1PI
kinematics is $x = s_4/m^2$ and the cross sections are
functions of $t_1$ and $u_1$.
In PIM kinematics the pair is treated as a unit so that, on the partonic level,
we have
\begin{eqnarray}
\label{qq_PIM}
i(k_1) + j(k_2) &\longrightarrow& Q \overline Q(p') +
X(k')\, . 
\end{eqnarray}
The square of the heavy quark pair mass is $p'^2 = M^2$.
At partonic threshold,
$X(k') = 0$, the three Mandelstam invariants are
\begin{eqnarray}
\label{tupidef}
s = M^2 \, ,\quad  t_1 = - \frac{M^2}{2} ( 1 - \beta_M\, \cos \theta )
\, ,\quad u_1 = - \frac{M^2}{2} ( 1 + \beta_M\, \cos \theta )\, 
\end{eqnarray}
where $\beta_M=\sqrt{1-4m^2/M^2}$ and $\theta$ is the scattering
angle in the parton center of mass frame.  Now the distance from threshold
is $x = 1 - M^2/s \equiv 1-z$ where $z = 1$ at threshold. In PIM kinematics
the cross sections are functions of $M^2$ and $\cos \theta$.

The resummation is done in moment space by making a Laplace
transformation with respect to $x$, the distance from threshold.  Then the
singular functions become linear combinations of $\ln^k \tilde{N}$ with
$k \leq l+1$ and $\tilde{N} = Ne^{\gamma_E}$ where $\gamma_E$ is the Euler
constant.  The 1PI resummed double differential partonic
cross section in moment space is 
\begin{eqnarray}
\label{resum:eq:9}
&& \hspace{-5mm}
s^2 \frac{d^2 \sigma^{\rm res}_{ij}(\tilde{N})}
{dt_1\,du_1}
\,=\,{\rm Tr}\Bigg\{
H_{ij} 
{\rm \bar{P}} \exp\left[\int\limits_{m}^{m/N} {d\mu'\over\mu'} 
(\Gamma^{ij}_S\left(\alpha_s(\mu^{\prime})\right))^{\dagger}\right] \!
{\tilde S}_{ij}
{\rm P} \exp\left[\int\limits_{m}^{m/N} {d\mu'\over\mu'} 
\Gamma^{ij}_S\left(\alpha_s(\mu^{\prime})\right)\right] \! \Bigg\} \\
\nonumber & & \times
\exp\left(\tilde{E}_{i}(N_u,\mu,\mu_R)\right)\, 
\exp\left(\tilde{E}_{j}(N_t,\mu,\mu_R)\right)\,\, \exp\Bigg\{ 2\,
\int\limits_{\mu_R}^{m}{d\mu'\over\mu'}\,\, \Bigl(
\gamma_i\left(\alpha_s(\mu^{\prime})\right) +
\gamma_j\left(\alpha_s(\mu^{\prime})\right) \Bigr) \Bigg\}\, ,
\end{eqnarray}
where $N_u=N(-u_1/m^2)$, $N_t=N(-t_1/m^2)$, 
and $(\overline {\rm P})$ P refer to
(anti-)path ordering.
To find the PIM result, transform $t_1$ and $u_1$ to $M^2$ and $\cos \theta$ 
using Eq.~(\ref{tupidef}) and let $N_u = N_t = N$. 
The cross section depends on the `hard',
$H_{ij}$, and `soft', $\tilde{S}_{ij}$, functions which are Hermitian matrices
in the space of color exchanges.  The `hard' part
contains no singular functions.  The `soft' component contains the singular 
functions associated with non-collinear soft-gluon emission.  
The soft anomalous
dimension matrix, $\Gamma_S^{ij}$, is two dimensional for $q \overline q$ and 
three for $gg$.
The universal Sudakov factors are in the exponents $\tilde{E}_i$, expanded
as
\begin{eqnarray}
\label{resum:eq:8}
\exp(\tilde{E}_{i}(N_u,\mu,m))
\simeq 1 + \frac{\alpha_s}{\pi}\left(\sum_{k=0}^2 C^{i,(1)}_{k}
\ln^k(N_u)  \right)
+ \left(\frac{\alpha_s}{\pi}\right)^2\left(
\sum_{k=0}^4 C^{i,(2)}_{k}\ln^k(N_u)
\right)
+\ldots \, \, .
\end{eqnarray}
These exponents also contain the effects of the singular functions due to
soft-collinear radiation.
The coefficients $C^{i,(n)}_k$, as well as the detailed derivation of the 
resummed and finite-order cross sections, can be found in Ref.~\cite{KLMV}.
The momentum space cross sections to NNLO-NNLL are obtained by gathering terms
at ${\cal O}(\alpha_s^3)$ and ${\cal O}(\alpha_s^4)$, inverting the Laplace
transformation and matching the $N$-independent terms in $H_{ij}$ and
$\tilde{S}_{ij}$ to the exact ${\cal O}(\alpha_s^3)$ results in 
Refs. \cite{NLOgg,NLOqq}.

Any difference in the integrated cross
sections due to kinematics choice arises from the ambiguity of the estimates.
At leading order the threshold
condition is exact and there is no difference between the total cross
sections in the two kinematic schemes. However, beyond LO 
additional soft partons are produced and 
there is a difference.  To simplify the argument, 
the total partonic 
cross section may be expressed in terms of dimensionless scaling functions
$f^{(k,l)}_{ij}$ that depend only on $\eta = s/4m^2 - 1$ \cite{KLMV},
\begin{eqnarray}
\label{scalingfunctions}
\sigma_{ij}(s,m^2,\mu^2) = \frac{\alpha^2_s(\mu)}{m^2}
\sum\limits_{k=0}^{\infty} \,\, \left( 4 \pi \alpha_s(\mu) \right)^k
\sum\limits_{l=0}^k \,\, f^{(k,l)}_{ij}(\eta) \,\,
\ln^l\left(\frac{\mu^2}{m^2}\right) \, .
\end{eqnarray} 
We have constructed LL, NLL, and NNLL approximations to $f_{ij}^{(k,l)}$ 
in the $q
\overline q$ and $gg$ channels for $k \leq 2$, $l \leq k$.  Exact results are
known for $k=1$ and can be derived using renormalization group methods for 
$k=2$, $l=1,2$ \cite{KLMV}.  
Our calculations use the exact LO and NLO cross sections with the
approximate NNLO-NNLL corrections.

The inclusive hadronic cross section is obtained by convoluting the
inclusive partonic cross sections with the parton luminosity $\Phi_{ij}$,
\begin{eqnarray}
\Phi_{ij}(\tau,\mu^2) &=& \tau \,\, \int\limits_{0}^{1}
dx_1\,\, \int\limits_{0}^{1} dx_2\,\, \delta(x_1x_2 - \tau)\,\,
\phi_{i/h_1}(x_1,\mu^2)\, \phi_{j/h_2}(x_2,\mu^2)\, ,
\label{parlum}
\end{eqnarray}
where $\phi_{i/h}(x,\mu^2)$ is the density of partons of flavor $i$ in
hadron $h$ carrying a fraction $x$ of the initial hadron momentum, at
factorization scale $\mu$.  Then
\begin{eqnarray}
\sigma_{h_1h_2}(S,m^2) &=& \sum\limits_{i,j = q,{\bar{q}},g} \,\,
\int\limits_{4m^2/S}^{1}\,\frac{d\tau}{\tau}\,\,\Phi_{ij}(\tau,\mu^2)\,\,
\sigma_{ij}(\tau S,m^2,\mu^2)\,  \label{sigtot} \\
&=& \sum\limits_{i,j = q,{\bar{q}},g} \,\,
\int_{-\infty}^{\log_{10}(S/4m^2-1)} d\log_{10}\eta \, \frac{\eta}{1+\eta} 
\ln(10) \, \Phi_{ij}(\eta,\mu^2)\,\,
\sigma_{ij}(\eta,m^2,\mu^2)\, \nonumber
\end{eqnarray}
where
\begin{eqnarray} 
\eta = \frac{s}{4 m^2} - 1\, = \, \frac{\tau S}{4 m^2} -1\, .
\label{eq:etadef}
\end{eqnarray}
Our investigations in Ref.~\cite{KLMV} showed
that the approximation should hold if the convolution of the parton densities 
is not very sensitive to the high $\eta$ region.

\begin{figure}[htb] 
\setlength{\epsfxsize=0.95\textwidth}
\setlength{\epsfysize=0.5\textheight}
\centerline{\epsffile{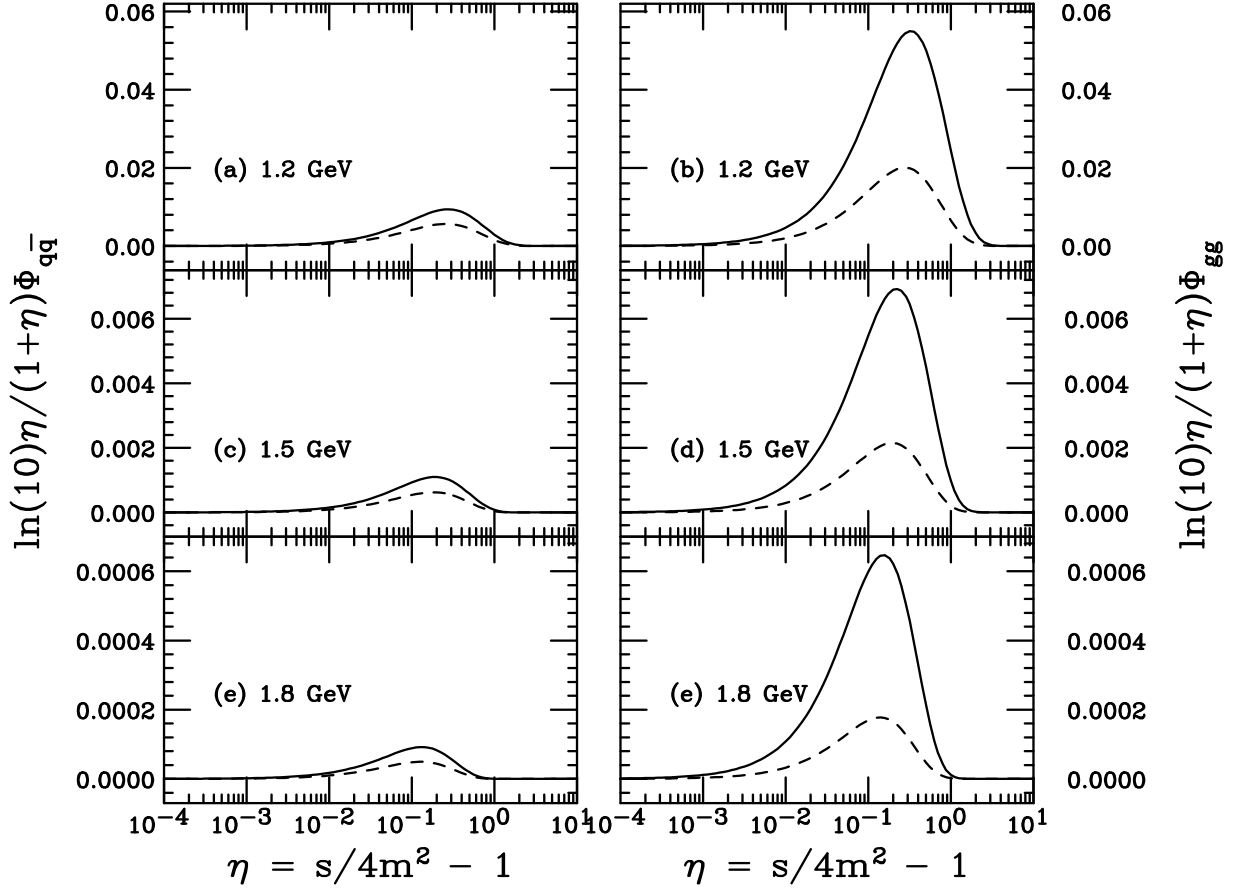}}
\caption[]{The parton luminosity for $pp$ interactions at $\sqrt{S} = 6.98$ GeV
as a function of $\eta$ using the GRV 98 HO densities.  
The left-hand side gives the $q \overline q$
luminosity, the right-hand side the $gg$ luminosity.  From top to bottom, the
charm quark mass is 1.2 GeV in (a) and (b), 1.5 GeV in (c) and (d), and 1.8 GeV
in (e) and (f).  The solid curves are with $\mu = m$ and the dashed, $\mu =
2m$. 
}
\label{pplum6} 
\end{figure}

\section{Numerical results}\label{sec:numerical}

In this section, we first test the applicability of our near-threshold 
treatment by calculating the parton luminosities.  We then compare our 
approximate NNLO-NNLL cross sections with the exact NLO cross sections and 
the $pp$ and $\pi^- p$ total charm cross sections.  We also discuss the 
convergence properties and scale dependence of our results.

\subsection{Parton luminosities}\label{subsec:lum}

\begin{figure}[htb] 
\setlength{\epsfxsize=0.95\textwidth}
\setlength{\epsfysize=0.5\textheight}
\centerline{\epsffile{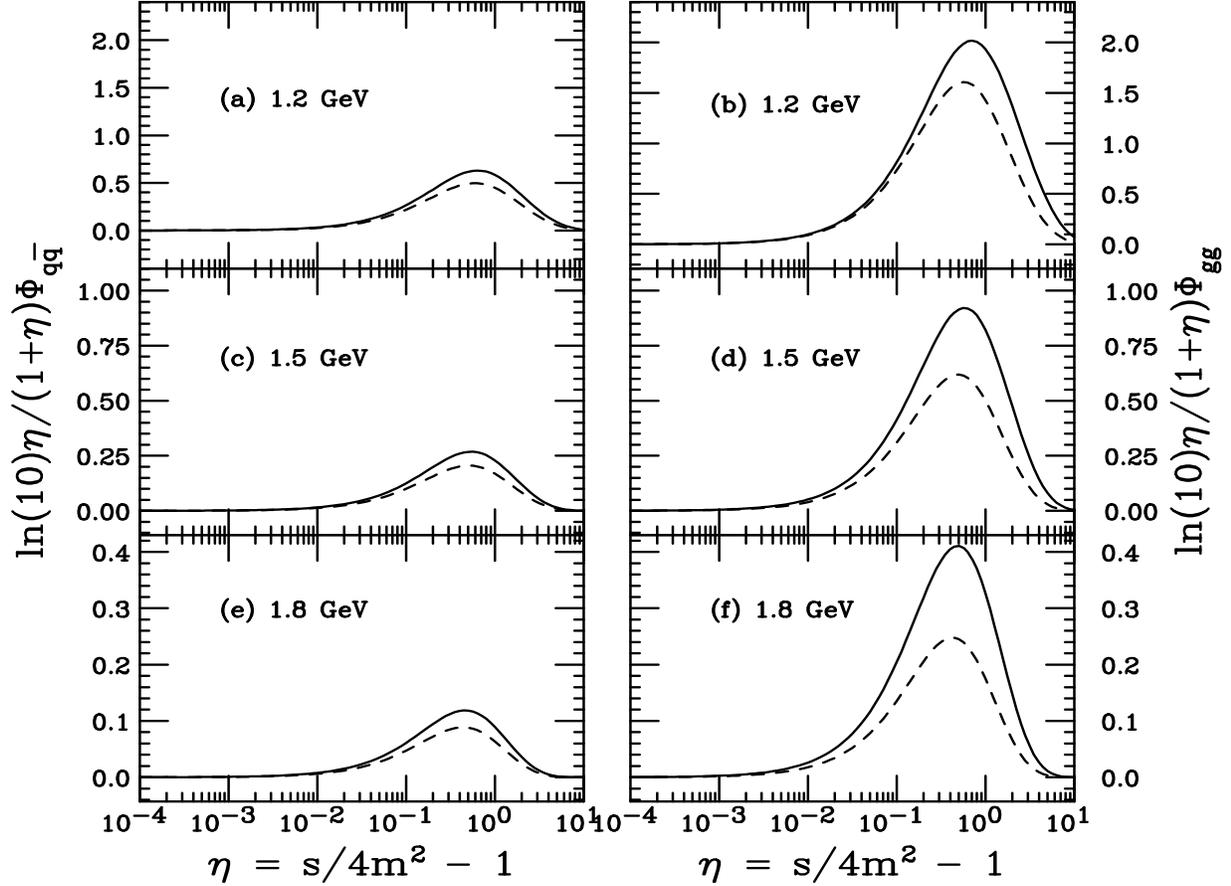}}
\caption[]{The parton luminosity for $pp$ interactions at $\sqrt{S} = 17.3$ GeV
as a function of $\eta$ using the GRV 98 HO densities.  
The left-hand side gives the $q \overline q$
luminosity, the right-hand side the $gg$ luminosity.  From top to bottom, the
charm quark mass is 1.2 GeV in (a) and (b), 1.5 GeV in (c) and (d), and 1.8 GeV
in (e) and (f).  The solid curves are with $\mu = m$ and the dashed, $\mu =
2m$. 
}
\label{pplum17} 
\end{figure}

\begin{figure}[htb] 
\setlength{\epsfxsize=0.95\textwidth}
\setlength{\epsfysize=0.5\textheight}
\centerline{\epsffile{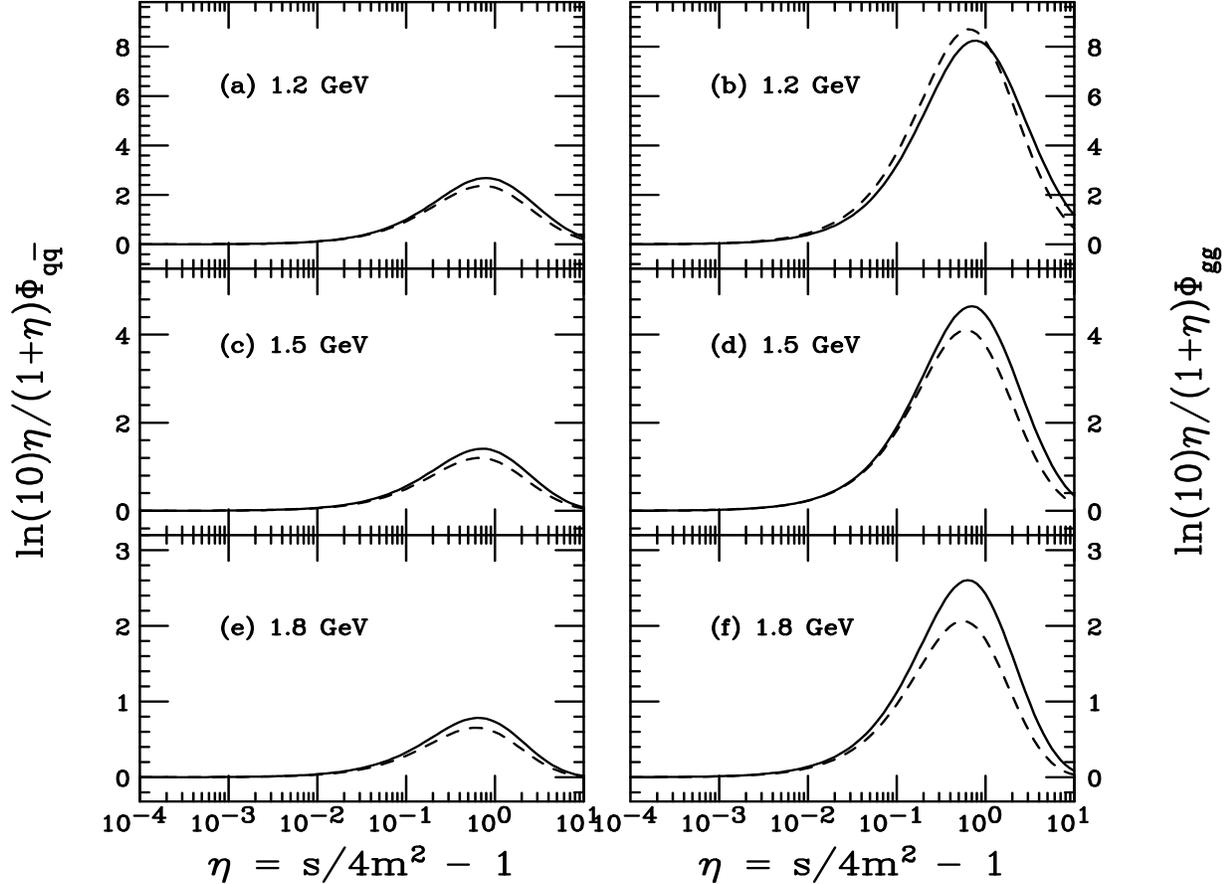}}
\caption[]{The parton luminosity for $pp$ interactions at $\sqrt{S} = 29.1$ GeV
as a function of $\eta$ using the GRV 98 HO densities.  
The left-hand side gives the $q \overline q$
luminosity, the right-hand side the $gg$ luminosity.  From top to bottom, the
charm quark mass is 1.2 GeV in (a) and (b), 1.5 GeV in (c) and (d), and 1.8 GeV
in (e) and (f).  The solid curves are with $\mu = m$ and the dashed, $\mu =
2m$. 
}
\label{pplum29} 
\end{figure}

We first check if the parton luminosity peaks at low enough values of $\eta$
for our near-threshold calculations to be applicable.  Figures
\ref{pplum6}-\ref{pplum29} show the parton luminosities for $\sqrt{S} = 6.98$,
17.3 and 29.1 GeV respectively, corresponding to the top
new GSI energy, the CERN
SPS lead beam, and the SPS proton beam.  The results are shown for three 
values of the charm quark mass at each energy: $m = 1.2$, 1.5 and 1.8 GeV.
Using the lowest value of $m$ gives relatively good agreement with $pp$ data
at NLO when $\mu = 2m$ \cite{HPC1,HPC2,wwproc}.  
This mass also represents a lower bound on the
possible range of charm quark masses.  The charm quark pole mass is most 
commonly chosen as 1.5 GeV.  Finally, 1.8 GeV represents an upper bound on 
the charm mass.

The GRV 98 HO parton densities \cite{grv98}
are used in the luminosity calculations, shown with
scales $\mu = m$ and $2m$.
The scale is not decreased below $m$ because the minimum scale in the parton
densities is larger than $m/2$.
Such calculations would thus be of little value.

Our results are primarily shown for the GRV 98 HO parton densities since they
are consistent with the most recent pion parton densities by Gl\"{u}ck, Reya
and Schienbein \cite{GRSpi2}, denoted here as GRS2.  
We will also compare our $pp$ results with those using CTEQ5M
\cite{cteq5} since these densities were used in our previous calculations for 
heavier quarks \cite{KLMV}.

At $\sqrt{S} = 6.98$ GeV, Fig.~\ref{pplum6}, 
charm production is well within the 
threshold region.  There is no luminosity at $\eta > 2$  
for $m = 1.5$ and 1.8 GeV while $\eta = 3$ is the highest $\eta$ with nonzero
luminosity for $m =
1.2$ GeV.  It is interesting to note that at even this rather low energy,
the $gg$ luminosity is still the largest, nearly a factor of ten greater than
the $q \overline q$ for $\mu = m$.  
The low $q \overline q$ luminosity can be attributed to
the steeply falling antiquark density at large $x$, $\mu/\sqrt{S} \sim 0.2- 
0.6$, over the charm mass range considered here.  
Each step in $m$ drops the luminosity by an order of 
magnitude while moving the peak to lower $\eta$, from $\eta
\sim 0.35$ when $m = 1.2$ GeV to $\sim 0.15$ when $m = 1.8$ GeV.  
Increasing the scale decreases the $gg$ luminosity by a factor of $\approx 3$.

The luminosity is considerably higher at $\sqrt{S} = 17.3$ GeV, 
the Pb+Pb center of mass
energy at the CERN SPS, as shown in Fig.~\ref{pplum17}. 
Now the peak in the luminosity is at higher $\eta$ but
still at $\eta < 1$, even for $m = 1.2$ GeV.  Thus, charm production may still
be considered as within the threshold region.  The location of the peak 
in $\eta$ is similar to that for $t \overline t$ production at the Tevatron
\cite{KLMV}.  Therefore, our calculations should be applicable.
Changing the scale does not have
a strong effect on the luminosity at this higher energy.  The relative $gg$
to $q \overline q$ luminosity is smaller than at the lower energy.
The higher SPS proton center-of-mass energy, $\sqrt{S} = 29.1$ GeV,
leads to a luminosity peak at somewhat larger
$\eta$, but remains less than unity, see Fig.~\ref{pplum29}.  Since some weight
is given to the region $\eta \sim 10$, especially for the lowest $m$
considered, this energy is the upper limit at which our calculation
is applied.

While all the calculations shown in Figs.~\ref{pplum6}-\ref{pplum29} have been
made for the GRV 98 HO densities, the results for the CTEQ5M densities are
quite similar.  The luminosities in $\pi^- p$ interactions do differ however.
At $\sqrt{S} = 6.98$ GeV, the $q \overline q$ luminosity is greater 
than the $gg$
luminosity by a factor of $\sim 1.5$ for $m = 1.2$ GeV, increasing to nearly a
factor of 5 for $m = 1.8$ GeV.  By $\sqrt{S} = 17.3$ GeV, the situation has
changed and the $gg$ channel again dominates but only by a factor of 1.1-1.5.
The smallest difference corresponds to the largest charm mass.  Finally, at
$\sqrt{S} = 29.1$ GeV, the $\pi^- p$ and $pp$ luminosities are rather similar.

The scaling functions that contribute to the partonic cross section
have been studied extensively in Ref.~\cite{KLMV}.  Since the scaling functions
are essentially independent of $m$, we do not show them again here.
We turn instead to a comparison of our calculations with the $pp$ and $\pi^- p$
total charm cross section data.

\subsection{Comparison with total charm data} \label{subsec:data}

Comparisons of the NLO cross sections to the available $c \overline c$ 
data have been made to obtain the `best' agreement by eye
with the data by varying the mass, $m$, and scale, $\mu$, for 
several combinations of $m$, $\mu$, and parton density \cite{HPC1,HPC2,wwproc}.
The best agreement with data at NLO is with $\mu = m =
1.3$ GeV for the GRV 98 densities and with $\mu =
m = 1.4$ GeV and $m = \mu/2 = 1.2$ GeV for CTEQ5M \cite{wwproc}.  
Thus the hadroproduction data tend to favor
rather light charm masses.

We now turn to our NNLO-NNLL results.  Since the NLO cross section is known
exactly, we add the ${\cal O}(\alpha_s^4)$ NNLL approximate contribution to 
the exact NLO cross section in the $q\overline q$ and $gg$ channels.
We also apply a damping factor, $1/\sqrt{1 +
\eta}$, to temper the influence of the approximate scaling functions at 
large $\eta$ where we are further from threshold and 
have less control over our approximations.

We find the greatest difference in the
kinematics schemes for $\mu = m$.  The effect of the NNLO-NNLL terms is reduced
for higher scales because $\alpha_s(m) > \alpha_s(2m)$.  The running of
$\alpha_s$ is significant for charm because the mass is rather low.  Thus the
strong running of the coupling constant for these values of $\mu$ ensures a
large scale dependence for charm production.  There is also a significant
parton density dependence in the results because $\mu/\Lambda_3$ is not large.
Since $\Lambda_3  = 0.38$ GeV for CTEQ5M and 0.2475 GeV for GRV 98 HO, the
NNLO-NNLL corrections will be larger for CTEQ5M than for GRV 98 HO.

The most important contribution to the NNLO-NNLL charm cross section is
$f_{gg}^{(2,0)}$, shown in Fig.~5 of Ref.~\cite{KLMV} for both 1PI and PIM
kinematics.  The differences between $f_{gg}^{(2,0)}$ in the two 
kinematics at large $\eta$ are considerable.  The functional dependence on 
$\eta$ begins to diverge for $\eta > 0.1$.  In 1PI kinematics, 
$f_{gg}^{(2,0)}$ is relatively small and positive until $\eta > 2$ when it
begins to grow.  On the other hand, in PIM kinematics $f_{gg}^{(2,0)}$ becomes
large and negative with increasing $\eta$.  This large negative contribution
can sometimes lead to a negative total cross section for $\mu =
m$ when the ${\cal O}(\alpha_s^4)$
contribution is larger than the NLO cross section.  
The decrease in $\alpha_s$ with
$\mu = 2m$ keeps the total PIM cross section positive even though the NNLO-NNLL
correction remains negative. 

We note here that there is some arbitrariness
in the functional form of the $gg$ scaling functions since the expression
$1-2t_1u_1/s^2$, used in the $gg$ Born cross section in Ref.~\cite{KLMV}, 
is equivalent at threshold to $(t_1^2+u_1^2)/s^2$.  Either may be used in the
scaling functions within the accuracy of our approximations.
However, different choices can lead to non-trivial numerical 
differences in the cross section.  We use the former choice here but point out 
that the latter choice increases the cross sections in both
kinematics and leads to less negative (or even positive) values for the
PIM results. 
We also note that including a class of subleading logarithms 
beyond NNLL, as derived in Ref. \cite{NK}, considerably lessens
the difference between the NNLO $q \overline q$ scaling functions in the two 
kinematics over a large $\eta$ region. For the $gg$ channel,
especially given the ambiguity in the functional form of the $gg$ scaling
functions described above, it is harder to draw firm conclusions. 
A full evaluation of all the subleading
terms requires two-loop calculations \cite{NK2} and has not yet been done.
However, the class of subleading logarithms arising from the 
inversion from moment to momentum space has been calculated exactly \cite{NK}.

As is shown below, the NNLO-NNLL cross sections in the 1PI and PIM kinematics 
are very different. Since including the subleading logs seems to make the 
scaling functions more similar, especially in the $q \overline q$ channel, 
we are motivated to present results using the average of the two kinematics 
as well as the individual 1PI and PIM results in the remainder of the paper.

\begin{figure}[htb] 
\setlength{\epsfxsize=0.95\textwidth}
\setlength{\epsfysize=0.5\textheight}
\centerline{\epsffile{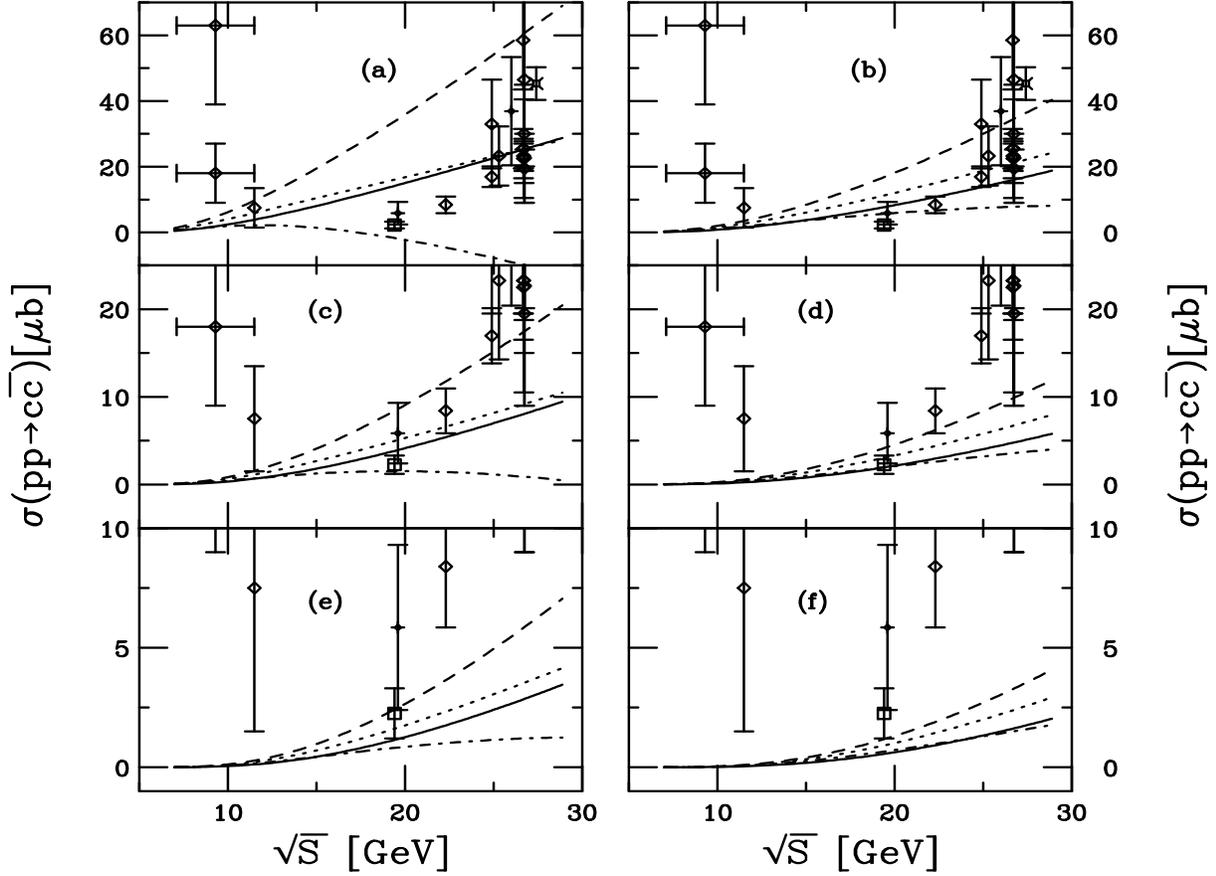}}
\caption[]{The total $c \overline c$ cross section in $pp$ interactions
as a function of $\sqrt{S}$ using the GRV 98 HO densities.  
The left-hand side employs the scale $\mu = m$, the right-hand side, $\mu =
2m$.  From top to bottom, the
charm quark mass is 1.2 GeV in (a) and (b), 1.5 GeV in (c) and (d), and 1.8 GeV
in (e) and (f).  The solid curves are the exact NLO result, the dashed curves,
the approximate 1PI NNLO-NNLL result, the dot-dashed 
curves, the approximate PIM NNLO-NNLL result, and the 
dotted curves, the average of the 1PI and PIM NNLO-NNLL results.
}
\label{ppgrv} 
\end{figure}

Our results with the GRV 98 HO densities are shown in Fig.~\ref{ppgrv} for $pp$
interactions with both
scales and the three charm quark masses.  The approximate NNLO-NNLL cross
sections are compared to the exact NLO results.  We plot the 1PI and PIM
cross sections as well as their average. The available data at
$\sqrt{S} < 30$ GeV are included.  In these figures, we are not attempting to
fit the data but to show the effect of the ${\cal O}(\alpha_s^4)$ correction.
The NNLO 1PI and PIM results with $\mu = m$ diverge most strongly from the 
NLO calculation. Increasing the charm mass reduces these differences, 
particularly for $\mu = 2m$.
The PIM correction at NNLO-NNLL is larger and negative, giving a negative 
total cross section for $\mu = m = 1.2$ GeV.

\begin{figure}[htb] 
\setlength{\epsfxsize=0.95\textwidth}
\setlength{\epsfysize=0.5\textheight}
\centerline{\epsffile{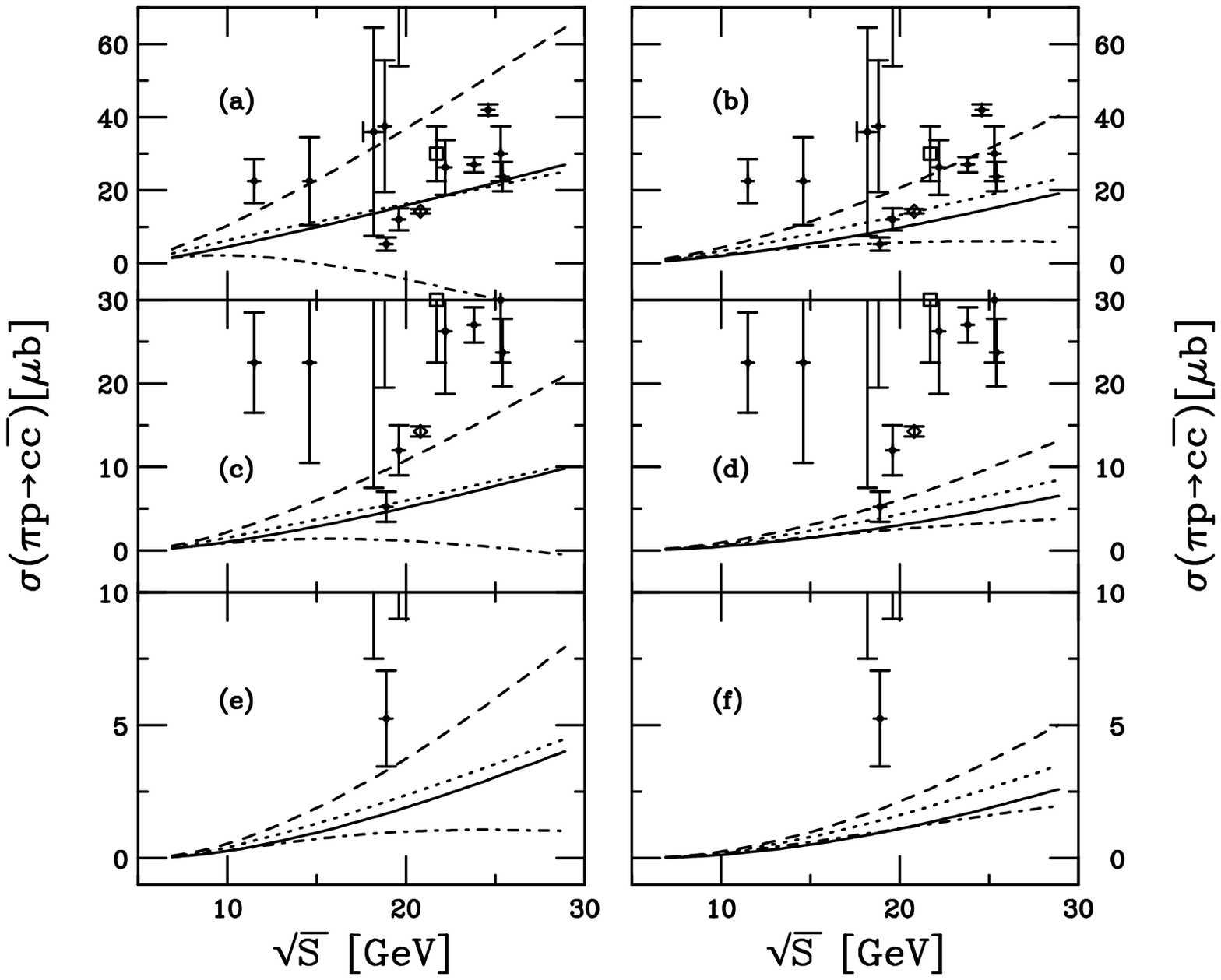}}
\caption[]{The total $c \overline c$ cross section in $\pi^- p$ interactions
as a function of $\sqrt{S}$ using the GRV 98 HO proton densities and the
GRS2 pion densities.  
The left-hand side employs the scale $\mu = m$, the right-hand side, $\mu =
2m$.  From top to bottom, the
charm quark mass is 1.2 GeV in (a) and (b), 1.5 GeV in (c) and (d), and 1.8 GeV
in (e) and (f).  The solid curves are the exact NLO result, the dashed curves,
the approximate 1PI NNLO-NNLL result, the dot-dashed 
curves, the approximate PIM NNLO-NNLL result, and the 
dotted curves, the average of the 1PI and PIM NNLO-NNLL results.
}
\label{pipgrv} 
\end{figure}

The same trends are shown in Fig.~\ref{pipgrv} for the GRV 98 HO-based 
calculations 
of the total cross section in $\pi^- p$ interactions.  We use the recently
updated pion parton densities, GRS2 \cite{GRSpi2}.  This evaluation has a
somewhat lower gluon density than the previous GRV-$\pi$ set
\cite{GRVpi}.

\begin{figure}[htb] 
\setlength{\epsfxsize=0.95\textwidth}
\setlength{\epsfysize=0.5\textheight}
\centerline{\epsffile{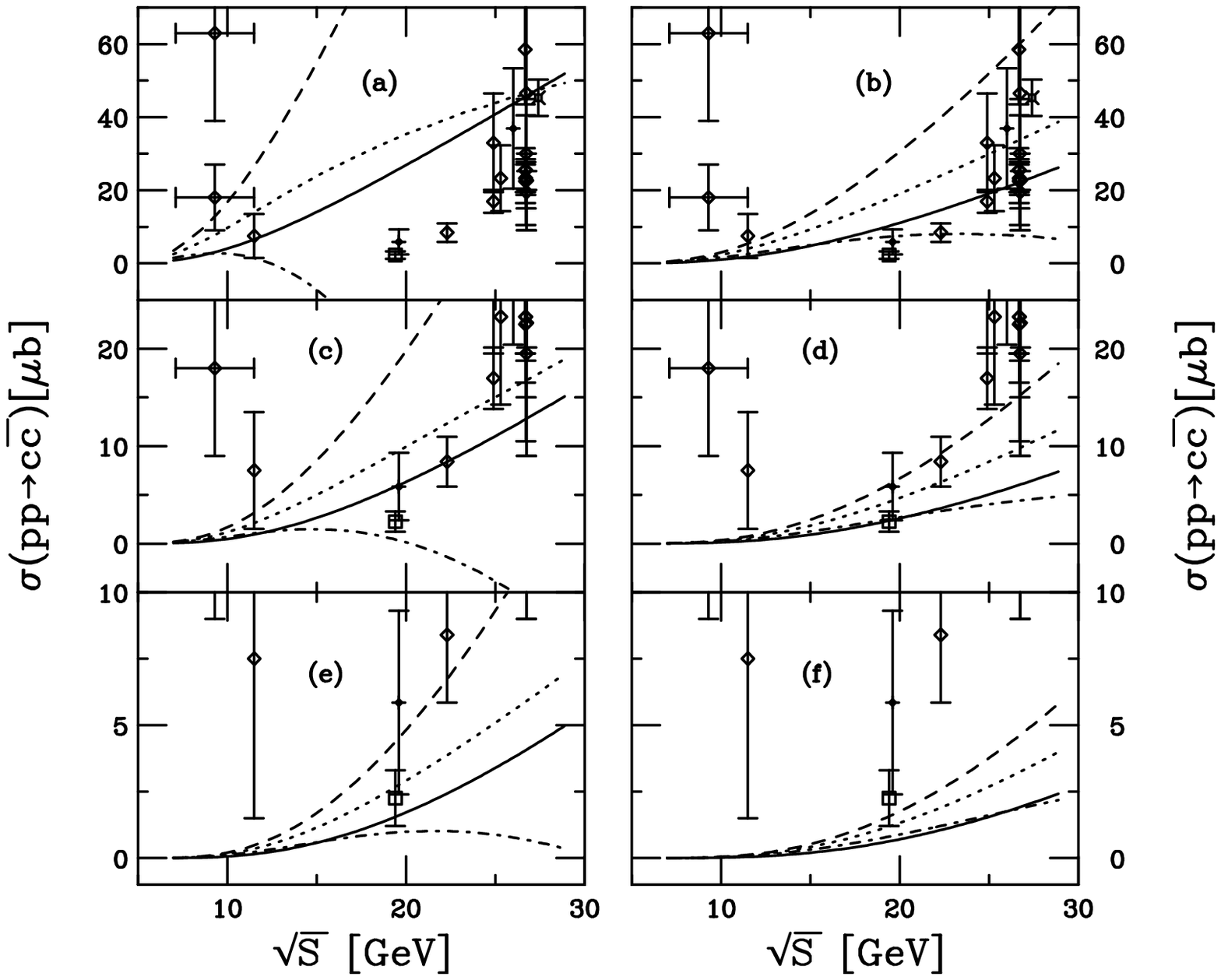}}
\caption[]{The total $c \overline c$ cross section in $pp$ interactions
as a function of $\sqrt{S}$ using the CTEQ5M densities.  
The left-hand side employs the scale $\mu = m$, the right-hand side, $\mu =
2m$.  From top to bottom, the
charm quark mass is 1.2 GeV in (a) and (b), 1.5 GeV in (c) and (d), and 1.8 GeV
in (e) and (f).  The solid curves are the exact NLO result, the dashed curves,
the approximate 1PI NNLO-NNLL result, the dot-dashed 
curves, the approximate PIM NNLO-NNLL result, and the 
dotted curves, the average of the 1PI and PIM NNLO-NNLL results.
}
\label{ppcteq} 
\end{figure}

We now show our $pp$ results with CTEQ5M in
Fig.~\ref{ppcteq}.  Since $\Lambda_3$ is larger for the CTEQ densities, 
all the cross sections are somewhat greater than those calculated 
with GRV 98 HO.   The larger $\Lambda_3$ results in a larger $\alpha_s$ and 
hence larger higher-order corrections. 
The NNLO 1PI and PIM results with $\mu = m$ diverge most strongly from the 
NLO calculation.  The PIM result for $m = 1.2$ GeV is already negative at 
$\sqrt{S} \sim 12$ GeV.  

\begin{table}[htb]
\begin{center}
\begin{tabular}{ccccccc} \hline
& \multicolumn{2}{c}{$\sigma({\rm NLO})$ ($\mu$b)} &
\multicolumn{2}{c}{$\sigma({\rm 1PI})$ ($\mu$b)} &
\multicolumn{2}{c}{$\sigma({\rm PIM})$ ($\mu$b)} \\
$\sqrt{S}$ (GeV) & $\mu = m$ & $\mu = 2m$ & $\mu = m$ & $\mu = 2m$ & 
$\mu = m$ & $\mu = 2m$ \\ \hline
\multicolumn{7}{c}{$m = 1.2$ GeV} \\ \hline
\multicolumn{7}{c}{GRV 98 HO} \\ \hline
6.98 & 0.45 & 0.11 & 1.29 & 0.29 & 0.70  & 0.21 \\
17.3 & 11.1 & 5.61 & 26.5 & 12.4 & 0.075 & 4.71 \\
29.1 & 28.8 & 18.8 & 69.0 & 40.3 & -12.9 & 8.09 \\ \hline
\multicolumn{7}{c}{CTEQ5M} \\ \hline
6.98 & 0.76 & 0.13 & 3.50  & 0.44 & 1.48   & 0.29 \\
17.3 & 19.5 & 7.33 & 75.5  & 20.5 & -17.0  & 6.32 \\
29.1 & 51.9 & 26.2 & 200.1 & 70.7 & -101.4 & 6.66 \\ \hline \hline
\multicolumn{7}{c}{$m = 1.5$ GeV} \\ \hline
\multicolumn{7}{c}{GRV 98 HO} \\ \hline
6.98 & 0.028 & 0.0062 & 0.079 & 0.017 & 0.056 & 0.014 \\
17.3 & 2.75  & 1.31   & 6.13  & 2.83  & 1.44  & 1.44  \\
29.1 & 9.44  & 5.77   & 20.4  & 11.9  & 0.47  & 4.00  \\ \hline
\multicolumn{7}{c}{CTEQ5M} \\ \hline
6.98 & 0.043 & 0.0077 & 0.17 & 0.025 & 0.11  & 0.019 \\
17.3 & 4.13  & 1.55   & 13.1 & 4.08  & 1.19  & 1.86  \\
29.1 & 15.1  & 7.37   & 46.3 & 18.5  & -8.50 & 4.82  \\ \hline \hline
\multicolumn{7}{c}{$m = 1.8$ GeV} \\ \hline
\multicolumn{7}{c}{GRV 98 HO} \\ \hline
6.98 & 0.0014 & 0.00030 & 0.0042 & 0.00086 & 0.0034 & 0.00074 \\
17.3 & 0.76   & 0.35    & 1.63   & 0.74    & 0.63   & 0.44    \\
29.1 & 3.45   & 2.13    & 7.05   & 4.08    & 1.24   & 1.77    \\ \hline
\multicolumn{7}{c}{CTEQ5M} \\ \hline
6.98 & 0.0024 & 0.00042 & 0.0094 & 0.0014 & 0.0070 & 0.0012 \\
17.3 & 1.01   & 0.38    & 2.89   & 0.97   & 0.82   & 0.54   \\
29.1 & 4.97   & 2.42    & 13.5   & 5.79   & 0.32   & 2.18   \\ \hline \hline
\end{tabular}
\caption{Charm total cross sections in $pp$ interactions
at the three fixed target energies
considered.  The results are given for the NLO exact, the 1PI and PIM NNLO-NNLL
approximate cross sections.  The $c \overline c$ cross sections using the 
GRV 98 HO and CTEQ5M parton densities are compared.}
\label{pptab}
\end{center}
\end{table}

The calculations agree only moderately well with the data.
The earliest data are all rather low statistics and are mostly measurements
of single $D$ mesons.  How the $c \overline c$ pairs hadronize is a 
particularly important question for energies near threshold where some channels
may be energetically disfavored.
We follow Ref.~\cite{mlm} and assume that since $\sigma(D_s)/\sigma(D^0 + 
D^+) \simeq 0.2$ and $\sigma(\Lambda_c)/\sigma(D^0 + D^+) \simeq 0.3$, it is
possible to obtain the total
$c \overline c$ cross section from $\sigma(D \overline D)$ by multiplying it by
$\approx 1.5$.  This assumption could have a strong energy dependence so that
as many charm hadrons as possible should be measured at each energy to
study hadronization.
The fragmentation into $D \overline D$, $D \overline \Lambda_c$, 
$\Lambda_c \overline D$, $\Lambda_c \overline \Lambda_c$,  {\it etc.}\  could
be studied at the GSI facility if the experiments are able to reconstruct charm
mesons and baryons.  

Some of the total cross section data are based on lepton measurements.
Semileptonic decays do not allow the momentum of the primary $D$ meson to be
entirely reconstructed, adding an additional layer of experimental uncertainty.
The CERN SPS results are based on muon spectrometers which cannot
unambiguously determine the identity of the primary hadron. Finally,
some of the data are taken on nuclear targets and then extrapolated to $pp$
assuming that the cross sections scale linearly with $A$, supported by 
fixed target measurements of the $A$ dependence of charm production \cite{alv}.
Thus the placing of data on the plots is primarily to guide the eye.

The total cross sections in $pp$ interactions
for the three energies we have discussed are 
given in Table \ref{pptab}.  The scale dependence is not
necessarily reduced at NNLO-NNLL relative to the exact NLO.  Due to the
complete dominance of the $gg$ channel in $pp$ interactions, the dependence
on kinematics choice is large.  Because $f_{gg}^{(2,0)}$ grows with $\eta$
at large $\eta$ in 1PI kinematics, the 1PI cross section is always larger 
than the NLO exact result.
On the other hand, the PIM result always underestimates the exact NLO
calculation since $f_{gg}^{(2,0)}$ is large and negative at large $\eta$
in PIM kinematics.  Thus the PIM cross section with $\mu = 2m$
can be greater than that with $\mu = m$.
The average of the 1PI and PIM cross sections typically remains somewhat
above the NLO result.

Recall that all the results include the damping factor.  
Without the damping factor, the
${\cal O}(\alpha_s^4)$ contribution can increase up to 40\% at the highest
energies.

Note that even for the NLO exact cross section, the CTEQ5M results are larger
than the GRV 98 HO results.  Because $\Lambda_3$ is greater for CTEQ5M, the
NNLO-NNLL corrections will be most important for this set.  The dependence on
parton densities is reduced for heavier 
quarks since the ratio $\mu/\Lambda \gg
1$ and a change in the ratio will not affect the value of $\alpha_s$
significantly for large $\mu$.  That is a primary reason why the
scale dependence is small for top quark production \cite{NK,KLMV}.  
In this case,
the running of $\alpha_s$ is primarily responsible for the dependence of the
results on parton density and the large kinematics dependence.

As discussed before, attempts to find agreement with the $pp$ and $\pi^- p$
charm production data by varying
the mass and scale usually result in a rather low value of $m$ at NLO, 
$\approx 1.2$ GeV for $\mu = 2m$.   The NNLO-NNLL results, particularly with
1PI kinematics, suggest that the same general energy dependence 
may be obtained for higher masses and lower scales so that a value of $m$
closer to the pole mass, 1.5 GeV, 
would agree with the fixed target data.  Using the GRV 98 HO 
densities, the NLO calculation with $m = 1.2$ GeV and $\mu = 2m$ is in
reasonable agreement with the approximate NNLO-NNLL 1PI result obtained with
$\mu = m = 1.5$ GeV.  Even at NNLO-NNLL, $m = 1.8$ GeV still underestimates the
data considerably.

Finally, we briefly compare our finite order result to the NLL-resummed result
for $pp \rightarrow c \overline c$ shown in Fig.~15 of Ref.~\cite{Bonci}.  Note
that the resummed cross section depends on the prescription used to obtain the
momentum space result.  Our NLO calculations with $m = 1.5$ GeV and CTEQ5M are
in good agreement with their NLO calculation using MRSR2 \cite{MRS2} since
these two sets have similar values of $\Lambda$.  The NLL resummed cross
section, calculated using the partonic total cross sections, is in rather good
agreement with our kinematics-averaged NNLO-NNLL results.

\subsection{Convergence properties} \label{subsec:kfac}

Given the large corrections at NNLO-NNLL, one can question if the calculated
cross section will stabilize such that all data can be described with one value
of the charm mass even if NNNLO and higher order corrections were
calculated. Since the NNLO-NNLL 1PI results allow the data to be described
relatively well with $\mu = m = 1.5$ GeV instead of $m = 1.2$ GeV and 
$\mu = 2m$,
it is quite possible that further corrections could lead to agreement with the
data for still higher values of $m$.  We can at least partially address this
issue through an investigation of the $K$ factors.

We first briefly discuss the calculation of the `first order $K$ factor',
$K^{(1)} = \sigma_{\rm NLO}/\sigma_{\rm LO}$.  Since $\sigma_{\rm LO}$ can
be calculated with LO or NLO parton densities, the value of the $K$ factor
depends on which way $\sigma_{\rm LO}$ is calculated.  In Ref.~\cite{RVkfac},
$K^{(1)}$ was defined in three ways.  We compare two of the definitions here:
$K_0^{(1)}$, where $\sigma_{\rm LO}$ is calculated with NLO parton
densities and a two-loop evaluation of $\alpha_s$, and $K_2^{(1)}$, where 
$\sigma_{\rm LO}$ is calculated with LO parton densities and a one-loop
evaluation of $\alpha_s$.  In both cases, the Born and ${\cal O}(\alpha_s^3)$
contributions to $\sigma_{\rm NLO}$ are calculated with NLO parton densities
and two-loop evaluations of $\alpha_s$.  The first definition, $K_0^{(1)}$,
indicates the convergence of terms in a fixed-order calculation while the
second, $K_2^{(1)}$, indicates the convergence of the hadronic calculation
towards a result.  If $K_0^{(1)} > K_2^{(1)}$, the convergence of the hadronic
cross section is more likely.

\begin{figure}[htb] 
\setlength{\epsfxsize=0.95\textwidth}
\setlength{\epsfysize=0.5\textheight}
\centerline{\epsffile{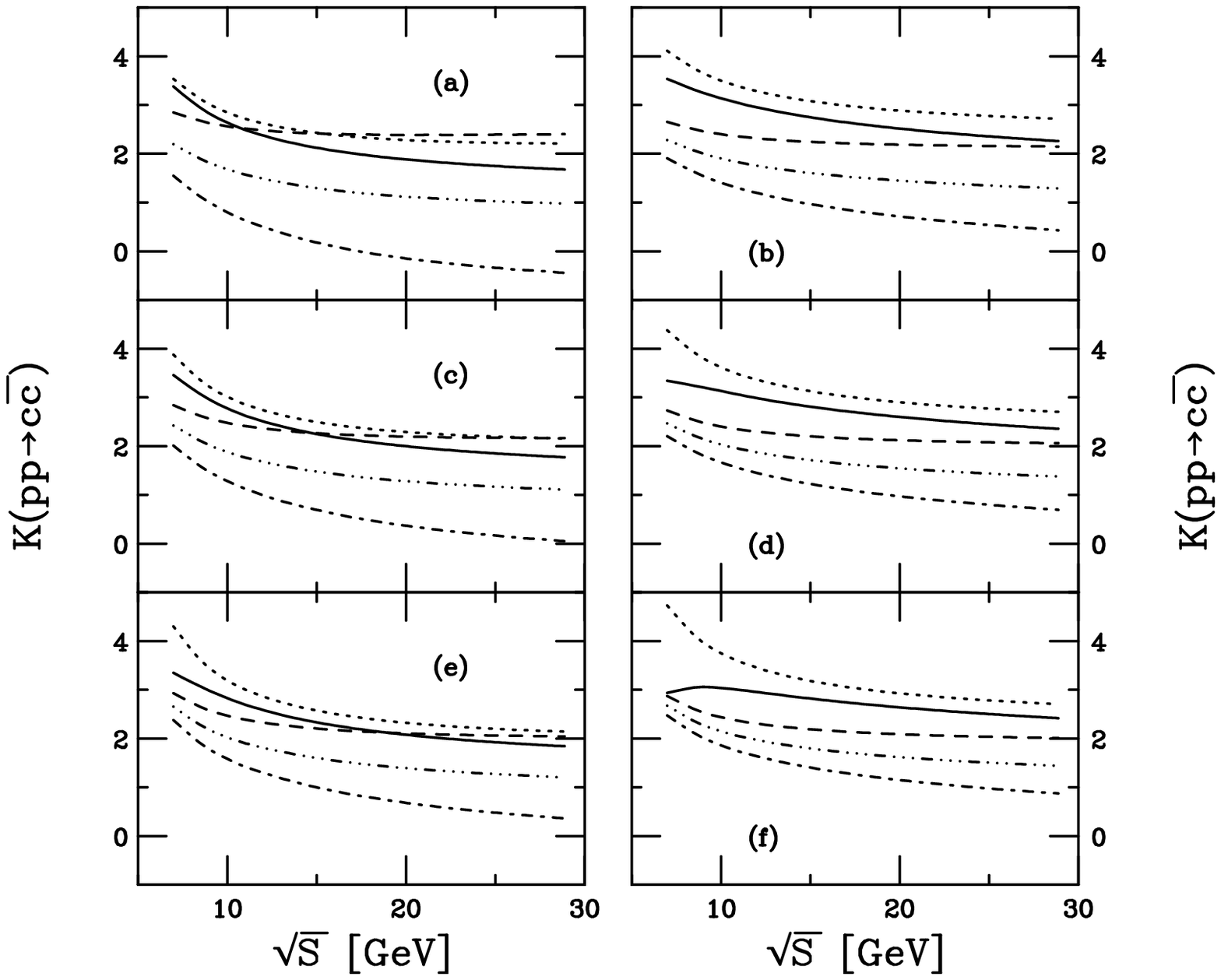}}
\caption[]{The theoretical $K$ factors for $c \overline c$ production in $pp$ 
interactions as a function of $\sqrt{S}$ using the GRV 98 HO densities.  
The left-hand side employs the scale $\mu = m$, the right-hand side, $\mu =
2m$.  From top to bottom, the
charm quark mass is 1.2 GeV in (a) and (b), 1.5 GeV in (c) and (d), and 1.8 GeV
in (e) and (f).  The solid curves are $K_2^{(1)}$, the dotted curves,
$K_0^{(1)}$, the dashed curves, $K_{\rm 1PI}^{(2)}$, the dot-dashed, 
$K_{\rm PIM}^{(2)}$, and the triple dot-dashed, $K_{\rm ave}^{(2)}$,
 as defined in the text.
}
\label{kfacgrv} 
\end{figure}

\begin{figure}[htb] 
\setlength{\epsfxsize=0.95\textwidth}
\setlength{\epsfysize=0.5\textheight}
\centerline{\epsffile{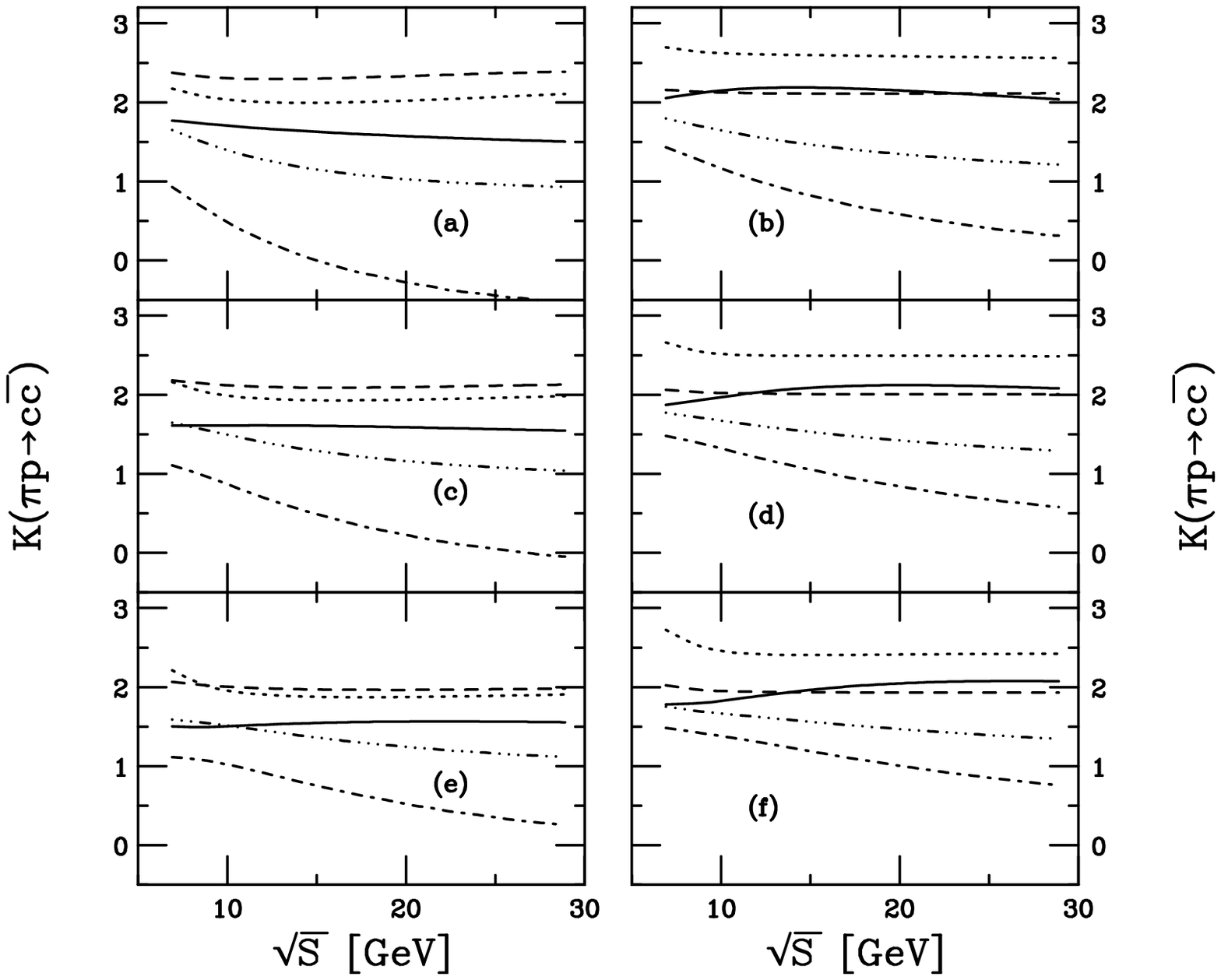}}
\caption[]{The theoretical $K$ factors for $c \overline c$ production in 
$\pi^- p$ interactions as a function of $\sqrt{S}$ using the GRV 98 
HO proton densities and the GRS2 pion densities.  
The left-hand side employs the scale $\mu = m$, the right-hand side, $\mu =
2m$.  From top to bottom, the
charm quark mass is 1.2 GeV in (a) and (b), 1.5 GeV in (c) and (d), and 1.8 GeV
in (e) and (f).  The solid curves are $K_2^{(1)}$, the dotted curves,
$K_0^{(1)}$, the dashed curves, $K_{\rm 1PI}^{(2)}$, the dot-dashed, 
$K_{\rm PIM}^{(2)}$, and the triple dot-dashed, $K_{\rm ave}^{(2)}$,
 as defined in the text.
}
\label{kfacpip} 
\end{figure}

In Fig.~\ref{kfacgrv}, we compare $K_0^{(1)}$ and $K_2^{(1)}$ for charm
production in $pp$ interactions with the GRV 98 HO densities.  The dotted
curves are $K_0^{(1)}$ and the solid curves are $K_2^{(1)}$.  Here, $K_0^{(1)}
> K_2^{(1)}$ for all values of $\mu$ and $m$ shown.  We see that for $\mu = m$,
$K_2^{(1)}$ does not change much with $m$.  It is $\approx 3.5$ for 
$\sqrt{S} = 6.98$ GeV, decreasing to $\approx
1.9$ at $\sqrt{S} = 29.1$ GeV for all cases. There is slightly more change with
$m$ for $\mu = 2m$ since $K_2^{(1)}$ tends to flatten at lower $\sqrt{S}$ with
increasing mass.  At higher $\sqrt{S}$, $K_2^{(1)}$ is somewhat bigger for 
the larger scale, $\approx 2.2$.  On the
other hand, $K_0^{(1)}$ is a stronger function of both $\mu$ and $m$.  When
$\mu = m$, $K_0^{(1)}$ increases from $\approx 3.7$ for $m=1.2$ GeV to 4.2 for
$m=1.8$ GeV at $\sqrt{S} = 6.98$ GeV, decreasing to $\approx 2.2$ for higher 
$\sqrt{S}$.  When $\mu = 2m$, $K_0^{(1)}$ increases 10-15\%.

If, instead, we use the CTEQ5M densities to calculate the NLO $K$ factors in
$pp$ interactions, $K_2^{(1)}>K_0^{(1)}$ at all energies and the values are all
larger than for the GRV 98 HO densities.  The larger $K$ factors
for CTEQ5M are due to the larger $\alpha_s$ with the CTEQ5M $\Lambda_3$.

More interesting are the $K$ factors in $\pi^- p$ interactions, shown
in Fig. ~\ref{kfacpip}.  With the GRV 98
HO densities, again $K_0^{(1)} > K_2^{(1)}$ for all $\mu$ and $m$ but the
values are much smaller and almost independent of energy.  When $\mu = m$,
$K_2^{(1)} \approx 1.5$ and $K_0^{(1)} \approx 2.2$ while when $\mu = 2m$,
$K_2^{(1)} \approx 2$ and $K_0^{(1)} \approx 2.5$.  Both the reduced size of
the $K$ factors as well as their near energy independence can be attributed to
the greater importance of $q \overline q$ annihilation in charm production,
particularly at low $\sqrt{S}$.  The ${\cal O}(\alpha_s^3)$ contribution to
the $q \overline q$ channel is considerably smaller relative to $\sigma_{\rm
LO}$ than in the $gg$ channel, reducing the $K$ factor.  Since the $pp$ cross
sections are dominated by the $gg$ channel, their $K$ factor is always large.

All these NLO $K$ factors are rather large for $pp$ interactions, typically
a factor of two or more.  We can define the next order $K$ factor, $K^{(2)} =
\sigma_{\rm NNLO-NNLL}/\sigma_{\rm NLO}$, and see if it is reduced relative to
$K^{(1)}$.  We define $K_{\rm 1PI}^{(2)}$, where the NNLO-NNLL 1PI cross 
section is in the numerator, $K_{\rm PIM}^{(2)}$, with the PIM cross 
section in the numerator, and $K_{\rm ave}^{(2)}$ where the numerator is the
average of the 1PI and PIM cross sections.  We cannot
distinguish between calculations with parton densities at different
orders for these $K$ factors, as we did for $K^{(1)}$, because there are no 
full NNLO parton densities available.  Now
both numerator and denominator are calculated with all NLO parton densities
and the two-loop evaluation of $\alpha_s$.  By calculating $K^{(2)}$ for the
NLO parton densities, we can identify improvements in the fixed-order results
relative to $K_0^{(1)}$.  

Figure~\ref{kfacgrv} also shows $K_{\rm 1PI}^{(2)}$, $K_{\rm PIM}^{(2)}$ and
$K_{\rm ave}^{(2)}$ in the dashed, dot-dashed and triple dot-dashed curves
respectively.  The 1PI $K$ factor, $K_{\rm 1PI}^{(2)}$, is not strongly
dependent on $\mu$, $m$ or $\sqrt{S}$.  It decreases slowly from $\approx 3$
at low $\sqrt{S}$ to $\approx 2$ at higher $\sqrt{S}$.  Consequently for $\mu =
m$, it is smaller than either calculation of $K^{(1)}$ at low $\sqrt{S}$ and
larger than $K^{(1)}$ as we move out of the threshold region.  When $\mu = 2m$,
$K_{\rm 1PI}^{(2)}$ is always less than $K^{(1)}$.  However, $K_{\rm
PIM}^{(2)}$ has the strongest $\mu$, $m$ and $\sqrt{S}$ dependence of all the
$K$ factors even though it is the smallest.  The fact that it is smaller is
less indicative of the ultimate convergence of the expansion than of
the fact that the NNLO-NNLL PIM contribution to the cross section is large and
negative.  This is demonstrated by the change of $K_{\rm PIM}^{(2)}$ from
positive to negative for $\sqrt{S} > 17$ GeV when $\mu = m = 1.2$ GeV.  As the
charm mass increases, the slope of $K_{\rm PIM}^{(2)}$ with $\sqrt{S}$
decreases and the $K$ factor remains positive.  
The average $K$ factor, $K^{(2)}_{\rm ave}$, is less than both $K_0^{(1)}$ and
$K_2^{(1)}$ everywhere.  It remains positive and is not strongly dependent on
$m$, $\mu$ or $\sqrt{S}$, remaining $\sim 2-3$ at low $\sqrt{S}$ and decreasing
to $\sim 1$ at higher $\sqrt{S}$.

Similar trends are observed for
the CTEQ5M densities but now the $K^{(2)}$ factors are all larger and more
dependent on $\mu$ and $m$, even $K_{\rm 1PI}^{(2)}$.  Now also $K_{\rm
PIM}^{(2)} < 0$ for $\mu = m = 1.2$ GeV and 1.5 GeV at large  $\sqrt{S}$.  
For $\pi^- p$ interactions, 
on the other hand, all the $K$ factors are again smaller 
and $K_{\rm 1PI}^{(2)}$ is almost independent of energy, as shown in 
Fig. \ref{kfacpip}.

We could form another $K$ factor, $K^{(2')} = \sigma_{\rm
NNLO-NNLL}/\sigma_{\rm LO} = K^{(2)} K_2^{(1)}$, to test the convergence of the
hadronic cross section.  The result is not complete because the NNLO-NNLL cross
section is only approximate and the NNLO parton densities are unavailable.
By multiplying the $K$ factors shown in Fig.~\ref{kfacgrv}, we see that $K_{
\rm 1PI}^{(2')}$ is $\approx 10$ at low $\sqrt{S}$ and $\approx 4-5$ at larger
$\sqrt{S}$.  Note that $K_{\rm ave}^{(2')}$ is smaller and $K_{\rm ave}^{(2')}
\sim K_2^{(1)}$ at high $\sqrt{S}$.
It is difficult to tell from these results if further, higher
order, $K$ factors such as $\sigma_{\rm NNNLO}/\sigma_{\rm NNLO-NNLL}$ will 
be consistently smaller than 2 near threshold or not.  The convergence of the
hadronic cross section at low scales is thus left in doubt although 
we note that subleading terms may have some effect on the convergence 
properties of the cross section \cite{NK}.  Since $K^{(2)} <
K_2^{(1)}$ for $\mu = 2m$, one can expect that the next-order correction might
be still smaller but this is not certain.  In any case, it is not a guarantee
that even larger masses may be needed to obtain the by-eye agreement at NLO
with further higher-order corrections.

\begin{figure}[htb] 
\setlength{\epsfxsize=0.95\textwidth}
\setlength{\epsfysize=0.5\textheight}
\centerline{\epsffile{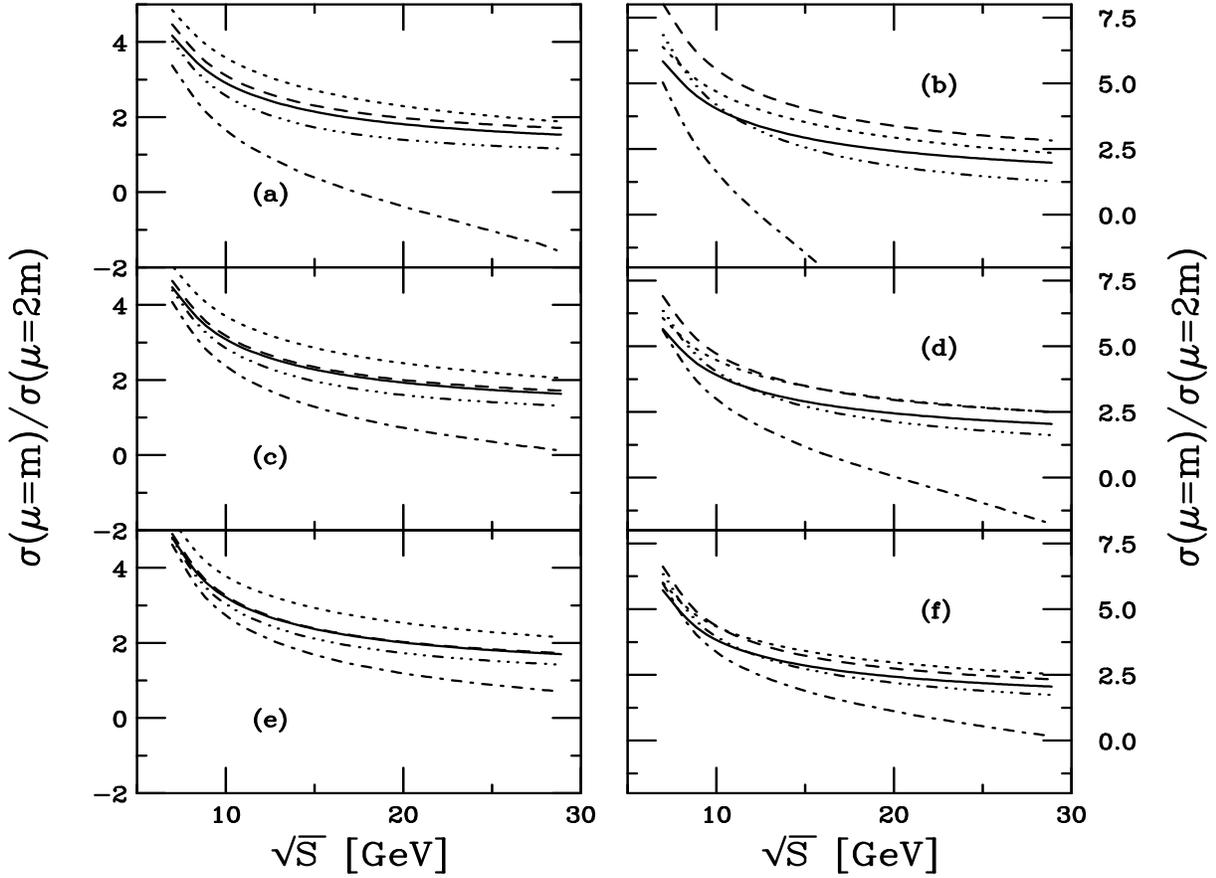}}
\caption[]{The scale dependence of $c \overline c$ production in $pp$ 
interactions as a function of $\sqrt{S}$.  
The left-hand side employs the GRV 98 HO parton densities, 
the right-hand side, CTEQ5M.  From top to bottom, the
charm quark mass is 1.2 GeV in (a) and (b), 1.5 GeV in (c) and (d), and 1.8 GeV
in (e) and (f).  The LO (dotted), NLO (solid), NNLO-NNLL 1PI (dashed), 
NNLO-NNLL PIM (dot-dashed) and the 1PI-PIM average (triple dot-dashed) 
ratios are shown.
}
\label{mudep} 
\end{figure}

\begin{figure}[htb] 
\setlength{\epsfxsize=0.55\textwidth}
\setlength{\epsfysize=0.5\textheight}
\centerline{\epsffile{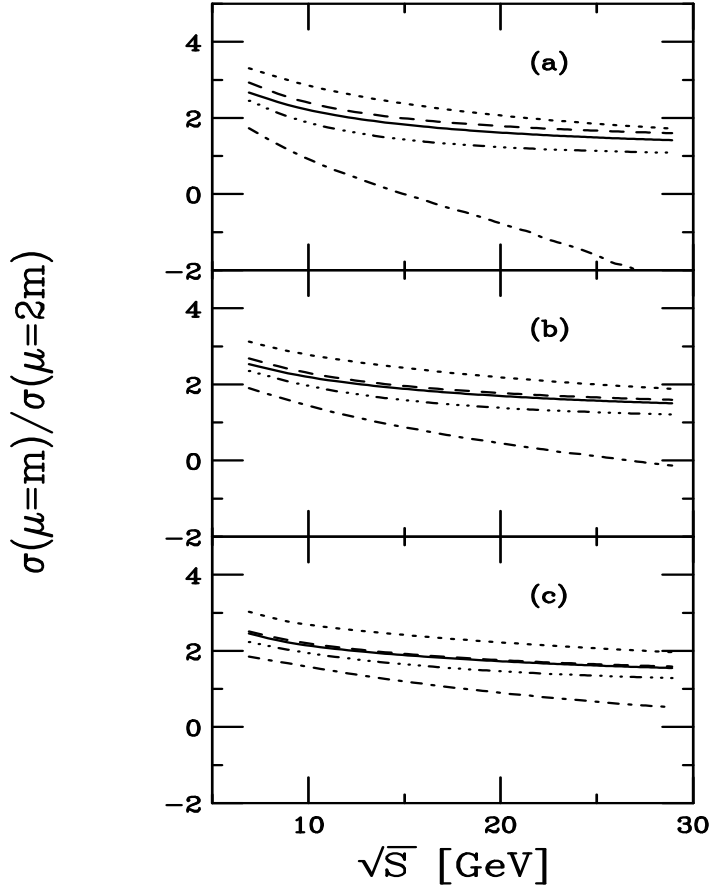}}
\caption[]{The scale dependence of $c \overline c$ production in $\pi^- p$ 
interactions as a function of $\sqrt{S}$, using the GRV 98 HO proton
densities and the GRS2 pion densities.  From top to bottom, the
charm quark mass is 1.2 GeV in (a) and (b), 1.5 GeV in (c) and (d), and 1.8 GeV
in (e) and (f).  The LO (dotted), NLO (solid), NNLO-NNLL 1PI (dashed), 
NNLO-NNLL PIM (dot-dashed) and the 1PI-PIM average (triple dot-dashed) 
ratios are shown.
}
\label{mudeppip} 
\end{figure}

\subsection{Scale dependence} \label{subsec:scale}

We now turn to a comparison of the scale dependence at LO, NLO and NNLO-NNLL,
shown in Fig.~\ref{mudep} for the ratio $\sigma(\mu = m)/\sigma(\mu = 2m)$ as a
function of $\sqrt{S}$ in $pp$ interactions.  
The GRV 98 HO scale dependence is shown on the
left-hand side and the CTEQ5M scale dependence on the right-hand side.
The GRV 98 HO scale dependence is largest for the LO ratio.  The NNLO-NNLL 1PI
ratio is somewhat larger than the NLO ratio for $m = 1.2$ GeV but the two are
almost identical for $m = 1.8$ GeV.  Thus for 1PI, the scale dependence is not
reduced relative to NLO but it is not really increased either.  The NNLO-NNLL
PIM ratio is smaller than the other ratios but it is a stronger function of
energy, again becoming negative for $m = 1.2$ GeV as $\sqrt{S}$ increases.
This ratio has the largest mass dependence.
The ratio for the average of the NNLO-NNLL 1PI and PIM cross sections 
remains below the NLO ratio for all energies shown.

The CTEQ5M scale dependence is considerably larger.  In this case, when $m =
1.2$ GeV, the NNLO-NNLL 1PI scale dependence is larger even than the LO,
dropping below it only when $m = 1.8$ GeV.  The overall increase in the scale
dependence relative to GRV 98 HO is again related to the larger $\alpha_s$ for
low scales with the CTEQ densities which also increases the NNLO-NNLL 
contribution to the total cross section.

The scale dependence of $\pi^- p$ production, shown in Fig. \ref{mudeppip},
is considerably reduced and, except for the
NNLO-NNLL PIM ratio, nearly independent of energy.  
This reduction in the scale
dependence can again be attributed to the relatively larger $q \overline q$
contribution to the total cross section.

\section{Conclusions}\label{sec:conclusions}

We have studied the behavior of  NNLO-NNLL calculations for charm production
near threshold in $pp$ and $\pi^- p$ interactions in both 1PI and PIM
kinematics. We find that there are large differences in the results
for different kinematics. Thus, the uncertainties in the cross sections  
remain large even at NNLO-NNLL. There are additional uncertainties, mentioned
earlier, due to the functional form of the scaling functions
and to subleading logarithms which have some effect on the values and
convergence of the 1PI and PIM cross sections.
We note that the average of the 1PI and PIM 
results has some nice properties such as better
convergence and less scale dependence.
For both $pp$ and $\pi^- p$ interactions, either choice of scale, and 
all values of $\sqrt{S}$ (except when $\mu=m=1.2$ GeV at high 
$\sqrt{S}$) , the average of the NNLO-NNLL 1PI and PIM results lies 
above the NLO cross section.
Thus, the charm mass need not be too low to agree with the data. 
However the poor convergence properties as well as other uncertainties
of the 1PI and PIM results make any quantitative statement about the
inclusive charm hadroproduction cross section difficult.

\subsection*{Acknowledgments}
The research of N.K. has been supported by a Marie Curie
Fellowship of the European Community programme ``Improving Human Research 
Potential'' under contract number HPMF-CT-2001-01221.
The work of E.L. is supported by the Foundation for 
Fundamental Research of Matter 
(FOM) and the National Organization for Scientific Research (NWO).
The work of R.V. is supported in part by the 
Division of Nuclear Physics of the Office of High Energy and Nuclear Physics
of the U.S. Department of Energy under Contract No. DE-AC-03-76SF00098.

\end{document}